# A New Quantum Ring Emitter of Anti-Whispering Gallery Modes

O'Dae Kwon, M. J. Kim, S.-J. An, S. E. Lee, D. K. Kim

*Department of Electronic and Electrical Engineering, Pohang University of Science and Technology, Pohang 790-784, Korea*
*E-mail: odkwon@postech.ac.kr*

**Abstract:** We have observed for the first time a new photonic quantum ring emission of anti-whispering gallery modes from a negative mesa-type toroid cavity due to semiconductor photonic corrals.

We have studied a photonic quantum ring (PQR) laser with naturally-formed quantum wire anomalies such as micro-to-nano ampere thresholds and $\sqrt{T}$ dependent spectral shifts, exhibited in the circumferential mesa region of active quantum well planes sandwiched by top and bottom DBR stacks [1], whose modes are optically confined by total internal reflections (TIRs) in a toroidal Rayleigh-Fabry-Perot cavity of 3 dimensional whispering gallery modes (WGM). Fig. 1 shows a CCD image of 15 μm PQR laser with a 10 μA injection current. The PQR generation is assumed to be associated with a photonic quantum corral effect (PQCE) in semiconductor [2, 3], and its hetero-spectral (multiple wavelength) radial emission characteristics are governed by an angular quantization rule [4].

We now report a new observation of anti-WGM type PQR emission from recent experiments on (1) hollow PQRs and (2) PQR holes(or negative-mesa): (1) For the hollow PQRs(Figs. 2), the inner boundary(hole) spectrum(curve a) was taken separately from the outer regular PQR spectrum(curve b) with a tapered fiber probe of 2 μm tip size. (2) Further, we also fabricated holes (or negative-mesas, Figs. 3(a)) and a typical emission spectrum is shown in Figs. 3(b) with changing view angles. We know that the old Rayleigh's WGM theory applies only to the 'concave surface' TIR. In contrast, the present hole emissions surprisingly arise from the convex surface, defying any common TIR knowledge. We thus believe a PQR type hole emission is generated here due to the strong the PQCE phenomenon, also assisted by additional gain guiding thereof.

We further fabricated an 8x8 PQR hole array in the same fashion. They exhibit fairly uniform emissions while the peak intensities gradually decrease as a function of distance from the p-electrode stripe. Present simple fabrication procedures with ease suggest an excellent potential application for high density optical interconnects as well.

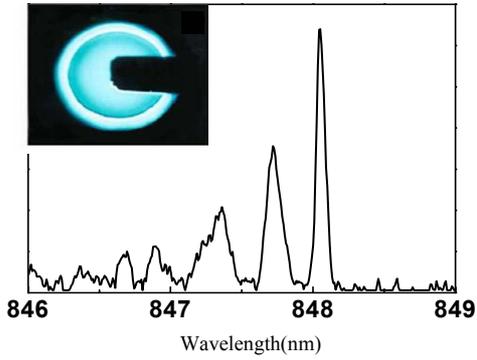
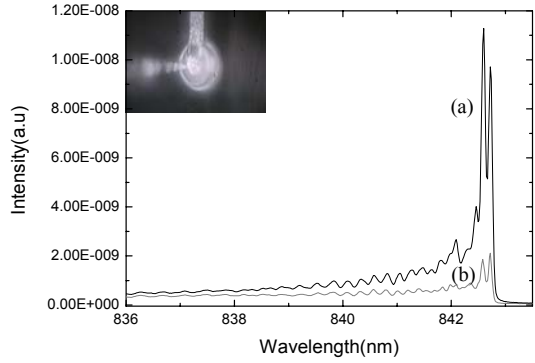

Fig. 1 PQR laser (diameter, $\phi$ =10 μm) spectrum measured in normal direction at I=800 μA. Spectral linewidth, FWHM= 0.055nm. The inset of a PQR laser CCD image of 15μm diameter at I=10 μA.

Fig. 2 Emission spectra from the inner hole ($\phi$=5 μm, solid curve a) and outer PQR ($\phi$=25 μm, dotted curve b) at I=3.9mA. The inset shows a CCD image of the hollow PQR with a tapered fiber probe at I=17 μA pointing to the inner hole

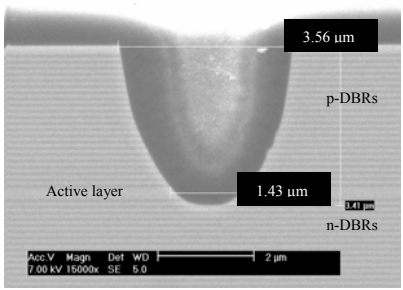
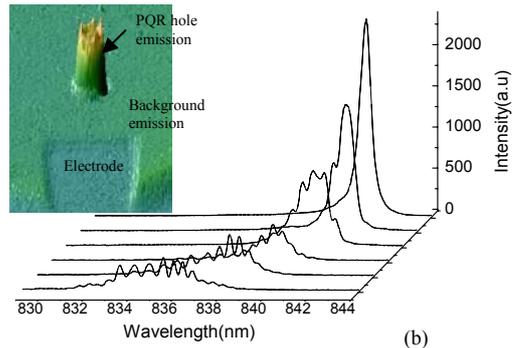

Fig. 3 (a) SEM image of a PQR hole cross section ($\phi$=1.43 μm). (b) angle-dependent emission spectra of a single PQR hole ($\phi$=6 μm) at I=100mA. (View angles are 0, 5, 10, 15, 20, 25° from the right, respectively) Inset shows a surface intensity profile emitted from the PQR hole at I=4mA.

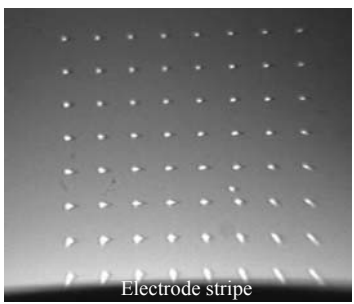
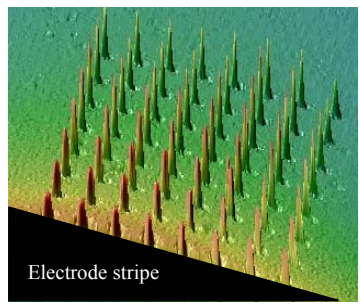

Fig. 4 (a) CCD image and (b) surface intensity profile from 8×8 array of 15 μm diameter PQR holes at I=10mA